\documentclass[journal]{IEEEtran}
\ifCLASSINFOpdf
\else
   \usepackage[dvips]{graphicx}
\fi
\usepackage[cmex10]{amsmath}
\begin{document}
%
\title{The Effects of Narrowband Interference on Finite-Resolution IR-UWB Digital Receiver}
%
%
%

\author{Chao Zhang,
        Huarui Yin  \itshape{Member}\upshape, 
        Pinyi Ren
\thanks{The work in this paper was supported in part by the National Science
Foundation of China under Grant No. 60802008 and the Oversea Academic
Training Funds (OATF), USTC.}
\thanks{C. Zhang and P. Ren are with the Department
of Information and Communication Engineering, Xi'an Jiaotong University, Shaan Xi, China,710049 P.R. China e-mail: \{chaozhang,pyren\}@mail.xjtu.edu.cn. C. Zhang was with the wireless information network lab (WINLab), University of Science and Technology of China.
}
\thanks{H. Yin is with the wireless information network lab (WINLab) and the
Department of Electronics Engineering and Information Science, University
of Science and Technology of China, Hefei, Anhui, 230027, P.R.China. Email:
yhr@ustc.edu.cn}
}

\maketitle

\begin{abstract}
Finite-resolution digital receiver is recently considered as a potential way to Ultra Wide Band (UWB) communication systems due to its ability of mitigating the challenge of Analog-Digital Converter (ADC). In this paper, the effects of narrowband interference (NBI) are investigated when finite-resolution digital receiver is used for Impulse Radio-UWB (IR-UWB) system. It is shown that  finite-resolution receiver enlarges the impact of NBI. The lower resolution of the UWB receiver is, the more degradations NBI causes.     
\end{abstract}


%
\IEEEpeerreviewmaketitle

\section{Introduction}
The potential strength of UWB system lies in its use of extremely wide bandwidth, which results in many attractive properties, e.g., high transmission rate and accurate position location \cite{IEEEhowto:coexist}. Due to its large transmission bandwidth, UWB systems need to coexist with a variety existing communication systems with relatively narrow bandwidth. Thus, these signals are called narrowband interference (NBI) signal in the view of UWB systems. The effect of NBI on various UWB systems based on analog matched filter were addressed by \cite{IEEEhowto:coexist} and \cite{IEEEhowto:lzhao}.\par

On the other hand, to implement digital UWB receiver, high sampling rate of analog-to-digital converter (ADC) is usually required for UWB signal \cite{IEEEhowto:mononit}\cite{IEEEhowto:FRreceiver}, and causes large challenges, e.g., unfordable power for high resolution ADC \cite{IEEEhowto:ADC}. 
For this purpose, finite-resolution digital UWB receiver with only one- or two-bit ADC, recently, were proposed by \cite{IEEEhowto:mononit}-\cite{IEEEhowto:monorev}. A common conclusion from their works is that full-resolution receiver are not recommended as its additional performance gains are too small to justify the increased implementation complexity. 
 To the best of our knowledge, the effect of NBI on finite-resolution UWB receiver has not been investigated yet.\par

  In this paper, we focus our attention on the effect of NBI on IR-UWB finite-resolution receiver. Two questions will be answered: 1) Whether does the finite-resolution receiver strengthen the harm resulted from NBI ? 2)  How to model the received symbol and evaluate the performance of finite-resolution receiver with NBI ?  To answer above two questions, we derive bit error ratio (BER) performances of both full- and finite-resolution receivers with NBI. Through analysis, we induce a linear signal model for finite-resolution receiver and answer that two questions. 
\section{System Model and Analysis }

To highlight the effect of NBI, we consider the reception
of a single-user scenario IR-UWB system. Assume $p_{tr}(t)$ denote the unit-energy transmitted pulse shape at IR-UWB transmitter, which incorporates the possible time-hopping sequence or direct-sequence spreading code if they are used. Denote $h(t)$ as the dispersive channel response function. We assume a slow varying channel. Let $p_{rec}(t)$ denote the  impulse response of low-pass filter (LPF) at the receiver with bandwidth $B$ and gain $1$. Then the received reference signal is 
\begin{equation}
w(t)=p_{tr}(t)\otimes h(t) \otimes p_{rec}(t)
\end{equation} 
where $\otimes$ denotes convolution.
%
Therefore, the filtered IR-UWB signal with NBI at the receiver can be expressed as
\begin{equation} \label{eq:eq1}
r(t)= \sqrt{E_s}\sum_{k=0}^\infty d_k w(t-kT)+r_I(t)+n(t)
\end{equation}
where $d_k$ is the $k$th transmitted symbol, which is equal to $\pm 1$
with equal probability, $r_I(t)$ is the filtered NBI signal, which could be a single-carrier. $E_s$ is the average transmission power and $n(t)$ is the Gaussian noise with zero-mean and variance $N_0/2$ per dimension. $T$ is the impulse period. For convenience, we rewrite (\ref{eq:eq1}) as
\begin{equation}
r(t)=r_0(t)+r_I(t)+n(t)
\end{equation}
where $r_0(t)$ is the desired received signal 
\begin{equation}
r_0(t)= \sqrt{E_s}\sum_{k=0}^\infty d_k w(t-kT)
\end{equation}
\subsection{Full-Resolution Receiver}
For a full-resolution receiver with channel state information, $r(t)$ will be fed into a matched filter. To reveal the effect of NBI, we only consider the receiver has no prior knowledge about the NBI. Hence, we employ a matched filter  as 
\begin{equation}
    {w_{mf}}(t) = w\left( {T - t} \right)
,~0\leq t \leq T 
\end{equation}
The output of matched filter is sampled at time $t=kT, k=0,1,\cdots$. Thus, for the $k$th symbol, the sampled signal is
\begin{equation}
    y[k]= \sqrt{E_s}E_w d_k+d_I[k]+n[k]
\end{equation}
where $y[k]=\int_{(k-1)T}^{kT}w(\tau)r(\tau)d\tau$, $E_w=\int_{(k-1)T}^{kT} w(\tau)^2d\tau $, $d_I[k]=\int_{(k-1)T}^{kT}w(\tau)r_I(\tau)d\tau$ and $n[k]=\int_{(k-1)T}^{kT}w(\tau)n(\tau)d\tau$. Without loss of generality, we set $E_w=1$. As $d_k$ is equal to $\pm 1$ with equal probability, then the bit error rate (BER) conditioned on $d_I[k]$ is 
\begin{equation}\label{eq:ber}
\begin{split}
   P_{mf}= \frac{1}{2}&\left[\mathcal{Q}\left( {\frac{{\sqrt{E_s} + {d_I}}}{{\sqrt {{N_0/2}} }}} \right) + \mathcal{Q}\left( {\frac{{\sqrt{E_s} - {d_I}}}{{\sqrt {{N_0/2}} }}} \right) \right]
\end{split}
\end{equation}
where we omit the subscript $k$ for convenience and $\mathcal{Q}(x)=\int_x^{+\infty}\frac{1}{\sqrt{2\pi}}e^{-\frac{t^2}{2}}dt$.
\subsection{Finite-Resolution Receiver}
For a finite-resolution digital receiver, the filtered signal is then
sampled at Nyquist rate $T_s = 1/(2B)$, and quantized to $b$-bit resolution. Herein, $b$ is usually less than 4 \cite{IEEEhowto:FRreceiver}. Yin \itshape{et al}\upshape \cite{IEEEhowto:mononit}  provided a linear Maximum Likelihood (ML) receiver, which is proved to be the optimal receiver of finite-resolution sampling\cite{IEEEhowto:FRreceiver}\cite{IEEEhowto:mononit}, to demodulate the sampled signals. Unfortunately, it is difficult to obtain the exact BER performance of the optimal finite-resolution receiver. However, \cite{IEEEhowto:mononit} also pointed out that a near-optimal finite-resolution  receiver built on the idea of Matched Filter, which is proposed in \cite{IEEEhowto:monorev}, has approximately equal demodulation weights to optimal receiver in low and middle SNR regime. For very high SNR regime, although the optimal receiver outperforms the near-optimal receiver, both BERs are usually far less than $10^{-6}$ which is usually reliable enough to data transmission. Thus the near-optimal receiver achieves nearly the same performance to the optimal receiver during our interested SNR range.  More  importantly, the error performance of near-optimal receiver is easier to be derived. Thus, we herein use the BER of near-optimal receiver to evaluate the effect of NBI in our interested case.

  Denote the b-bit quantized version of $r(t)$ as $\hat{r}(t)=\mbox{Q}_b(r(t))$. Set the output set of quantizer as $\{r_1,r_2,...,r_{2^b}\}$ and quantizer level as $ {q_1,...,q_{2^b-1}}$. Then $\mbox{Q}_b(r)=r_i$, if $q_{i-1}\leq r< q_{i}$. Since the quantization is a non-linear process, it is difficult to analyze the system performance. Thanks to the linearization method (Bussgang theorem) introduced in  \cite{IEEEhowto:linear} and \cite{IEEEhowto:OFDM} for a Gaussian input, we can describe $\hat{r}(t)$ as
\begin{equation}
    \hat{r}(t)=\alpha_b r(t) + v(t)
\end{equation}
where $\alpha_b=E\{\mbox{Q}_b'(r(t))\}$ is the linear gain and $v(t)$ is the nonlinear distortion and follows normal distribution with zero mean and variance $\sigma_{b}^2$. Assume optimal quantizer is employed, where quantization levels are properly chosen so that overall quantization error is minimized  (See \cite{IEEEhowto:optimal} to find optimal $\{r_i\}$ and $\{q_i\}$), therefore, $\alpha_b=\sum_i \int (q_i-q_{i-1}) \delta(r-q_i) f(r) dr$ where $f(r(t))$ is the probability density function of $r(t)$ and $\delta(x)$ is the Dirac delta function. So we also can obtain $\sigma_b^2=E\{\hat{r}^2(t)\}-\alpha_b^2E\{r^2(t)\}$. The near-optimal receiver provided in \cite{IEEEhowto:monorev} could be expressed as sgn$(\hat{r}(t)\otimes \omega(T-t))$ . Thanks to the linearization process,
the sampled signal after matched-filter is modeled as  
\begin{equation}
    \hat{y}[k]= \sqrt{E_s}\alpha_b d_k +\alpha_b d_I[k]+\alpha_b n[k] + v[k]
\end{equation}
where $v[k]=\int_{(k-1)T}^{kT}w(\tau)v(\tau)d\tau$ and  $\alpha_{b}=E\{\mbox{Q}'(r(kT))\}$. Follow the derivation of (\ref{eq:ber}), the BER of finite-resolution receiver conditioned on $d_I[k]$ is
\begin{equation}\label{eq:ber_fr}
\begin{split}
 P_{fr}=\frac{1}{2}&\left[\mathcal{Q}\left(\frac{ \sqrt{E_s}+d_I}{\sqrt{ N_0/2+\sigma_{b}^2/\alpha_b^2}}\right)\right.\\
&\left.+\mathcal{Q}\left(\frac{\sqrt{E_s}-d_I}{\sqrt{ N_0/2+ \sigma_{b}^2/\alpha_b^2}}\right)\right]
\end{split}
\end{equation}
Note that if we substitute $d_I$ with its absolute value $|d_I|$, the value of  $P_{fr}$ does not change. Therefore, we could analyze (\ref{eq:ber}) and (\ref{eq:ber_fr}) with $d_I \geq 0$ or $|d_I|$. If the UWB system is designed  for a desired reliability, e.g., $BER < 10^{-6}$, there should be $$\frac{ \sqrt{E_s}\pm d_I}{\sqrt{ N_0/2}} \gg 0~~\mbox{and}~~\frac{ \sqrt{E_s}\pm d_I}{\sqrt{ N_0/2+\sigma_{b}^2/\alpha_b^2}} \gg 0$$ with high probability during an impulse period. It is also to say the linearization method used in this paper works well if the correlation between $r_I(t)$ and $w(t)$ is very small, which could be assured by proper design of $p_{tr}(t)$, e.g., direct-sequence spreading code. 
In this case, $\mathcal{Q}(x)$ is a convex function for $x>0$. In following discussions, we thus only consider the reliable transmission case. 
\par
\itshape{\textbf{Remarks}}:\upshape
\begin{enumerate}
    \item From (\ref{eq:ber_fr}), we can see that if $\sigma_{b}^2/\alpha_b^2$ increases $P_{fr}$ increases. Results derived by \cite{IEEEhowto:linear} show that quantizer with more bits causes a less $\sigma_{b}^2/\alpha_b^2$. As a result, lower resolution (smaller b ) receiver incurs higher BER.  Also by  (\ref{eq:ber_fr}) and (\ref{eq:ber}), given  $\mathcal{Q}(x)$ is a convex function, we can deduce that $P_{mf}$ and $P_{fr}$ increase as $|d_I|$ increases. Therefore, both finite-resolution sampling and NBI can degrade the system performance. 

\item Also due to the property of convex function, we obtain \begin{equation*}
\begin{split}
 &P_1=P_{fr}(d_I\neq 0)-P_{fr}(d_I=0)>0\\
 &P_2=P_{fr}(d_I=0)-P_{mf}(d_I=0) >0   \\
 &P_3=P_{mf}(d_I\neq 0)-P_{mf}(d_I=0)>0 
\end{split}
\end{equation*}
Moreover, $P_1$ and $P_3$ increases as $|d_I|$ increases and $P_1$ and $P_2$ increase as $\sigma_{b}^2/\alpha_b^2$. Hence, $$P_0(d_I)=P_{fr}(d_I)-P_{mf}(d_I) >0$$ where $P_0$ increases as $|d_I|$ or $\sigma_{b}^2/\alpha_b^2$ increases. As a result, there is
$$P_0(d_I\neq 0)-P_0(d_I=0)>0$$ and the difference increases as  $\sigma_{b}^2/\alpha_b^2$ increases. That is to say lower resolution receiver with the same NBI causes extra degradation incurred by NBI. In other words, finite-resolution receiver strengthens the impact of NBI and NBI enlarges the performance gap between full-resolution and finite-resolution receivers.

    \item Compare (\ref{eq:ber}) and (\ref{eq:ber_fr}), we found the only difference is that there is an extra term $\sigma_{b}^2/\alpha_b^2$ in noise variance. Therefore, we can model the finite-resolution receiver as 
\begin{equation}\label{eq:linear}
    \hat{y}[k]= \sqrt{E_s}d_k+d_I[k]+m[k] 
\end{equation}
where $m[k]$ is the equivalent Gaussian noise  of finite-resolution receiver with zero-mean and variance $N_0/2+\sigma_{b}^2/\alpha_b^2$. The
linear signal model in (\ref{eq:linear}) provides a general method to analyze finite-resolution receiver. 
     
\end{enumerate}

\section{Numerical Results}
In this section, we verify our theoretical results and remarks. We consider a second derivative Gaussian pulse with time constant $\tau=0.16$ ns to meet regulation of UWB. We also assume there is no ISI and timing is perfect. The multi-path fading channel we used is the standard CM1 channel model\cite{IEEEhowto:cm1}. The simulation parameters are as
follows. The filter bandwidth was $B=8$ GHz. Noise variance is $N_0/2=1/2$. We consider 1-bit and 2-bit (3-level \cite{IEEEhowto:FRreceiver}) finite-resolution receivers in our simulations. According to the optimal quantizer proposed in \cite{IEEEhowto:optimal}, we can obtain:  $\alpha_1=0.7979$ and $\sigma_1^2=0.23$ for 1-bit receiver; $\alpha_2=0.8829$ and $\sigma_2^2=0.11$ for 2-bit receiver. To illustrate clearly, we define signal-noise ratio as SNR$=2E_s/N_0$ and signal-interference ratio as SIR$=E_s/E\{r_I(t)^2\}$. The carrier frequency of NBI is 5 GHz. We model the NBI as a BPSK modulation signal (a tone signal incurs the same performance due to \cite{IEEEhowto:coexist}).  Note that the simulation results in this paper  are obtained through intensive Monte Carlo experiments. As it is difficult to derive the distribution of $d_I$, we have to average (\ref{eq:ber}) and (\ref{eq:ber_fr}) over $10^8$ realizations of $d_I$ to calculate our theoretical results. \par

\begin{figure}[!t]
\centering
\includegraphics[width=3in]{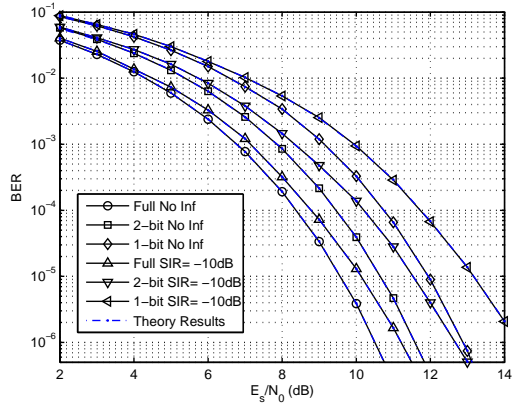}
\caption{BER performance of different receivers with SIR=-10dB}
\label{fig:fig1}
\end{figure}

Fig. 1 and Fig. 2 show  BER performances of different UWB receivers in a standard CM1 channel where NBI exits.
In both figures, \textquoteleft\textbf{Full}\textquoteright\ denotes full-resolution receiver, \textquoteleft\textbf{No Inf}\textquoteright\ means no interference, and \textquoteleft\textbf{n-bit}\textquoteright\ denotes finite-resolution receiver with n-bit sampling. First of all, we can see that our theoretical results fit the simulation curves closely for all cases. It verifies our analysis and the linear model. For a specific BER=$10^{-5}$ without NBI, full-resolution receiver can save about 0.8 dB and 2.3 dB SNR respectively, compared with 2-bit receiver and 1-bit finite-resolution receiver. Obviously, as SIR increases BER of each receiver increases. Thus the item 1) of remarks is verified. When SIR$=-10$ dB and BER=$10^{-6}$, full resolution receiver loses about 0.7 dB, 2-bit receiver loses about 1.2 dB and 1-bit receiver loses about 1.5 dB than its corresponding receiver without NBI respectively. We also can see from Fig. 2 that when SIR$= -15$dB and BER=$10^{-6}$, full-resolution receiver without NBI achieves about 2 dB gain over full-resolution with NBI, about 2.3 dB in the case of  2-bit receiver and about 3 dB in the case of 1-bit receiver. Through comparing, we find that finite-resolution receiver indeed enlarges the degradation incurred by NBI and lower resolution receiver causes greater enlargement.

\section{Conclusion}
In this paper, we provided BER performances of both full-resolution and finite-resolution IR-UWB receiver in the presence of NBI. We found that both finite-resolution sampling and interference can degrade the receiver performance and finite-resolution receiver strengthens the harm of NBI.


%

\appendices

\ifCLASSOPTIONcaptionsoff
  \newpage
\fi



%

\begin{figure}[!t]
\centering
\includegraphics[width=3in]{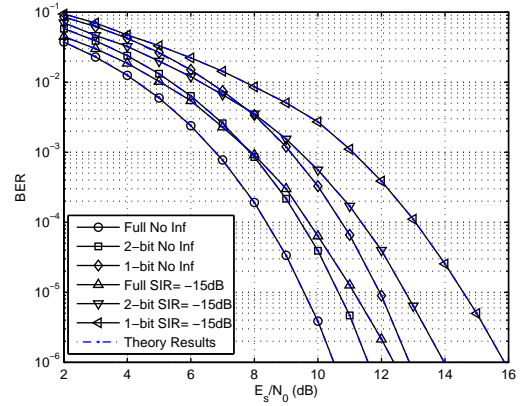}
\caption{BER performance of different receivers with SIR=-15dB}
\label{fig:fig2}
\end{figure}

%








\end{document}